\documentclass[amsmath,amssymb,aps,showkeys,reprint, %
floatfix,groupedaddress,superscriptaddress,amsmath,amssymb, %
prd]{revtex4-1}

\usepackage{dcolumn}
\usepackage{amsmath,amssymb}
\usepackage[breaklinks = true,colorlinks = true,%
linkcolor = blue,citecolor = blue]{hyperref}
\usepackage{graphicx}
\usepackage{xcolor}
\usepackage{ascmac}
\usepackage{slashed}

\newcommand{\diag}{\mathrm{diag}}

\newcommand{\bv}{\boldsymbol{v}}
\newcommand{\boldell}{\boldsymbol{\ell}}

\newcommand{\calC}{\mathcal{C}}
\newcommand{\calF}{\mathcal{F}}

\newcommand{\thetab}{\theta_{\rm B}}

\newcommand{\dd}{\langle dd \rangle}
\newcommand{\phiud}{\Phi_{\rm 2SC}}
\newcommand{\phidd}{\Phi_{dd}}
\newcommand{\deltadd}{\Delta_{dd}}

\newcommand{\qop}{\hat{q}}

\begin{document}

\title{Alice meets Boojums in neutron stars: \\
vortices penetrating two-flavor quark-hadron continuity}

\author{Yuki~Fujimoto}
\email{fujimoto@nt.phys.s.u-tokyo.ac.jp}
\affiliation{Department of Physics, The University of Tokyo,
  7-3-1 Hongo, Bunkyo-ku, Tokyo 113-0033, Japan}

\author{Muneto~Nitta}
\email{nitta@phys-h.keio.ac.jp}
\affiliation{Department of Physics \& Research and Education Center for Natural Sciences,
Keio University, Hiyoshi 4-1-1, Yokohama, Kanagawa 223-8521, Japan}

\begin{abstract}
  Alice and Boojums are both representative characters created by
  Lewis Carroll.  We show that they possibly meet in cores of rotating
  neutron stars.
  Recent studies of quark-hadron continuity suggest that neutron
  superfluid matter can connect smoothly to two-flavor symmetric quark
  matter at high densities.  We study how this can be maintained in the
  presence of the vortices.
  In the neutron matter, quantized superfluid vortices arise.  In the
  two-flavor dense quark matter, vortices carrying color magnetic
  fluxes together with fractionally quantized superfluid circulations
  appear as the most stable configuration, and we call these as the
  non-Abelian Alice strings.
  We show that three integer neutron superfluid vortices and three
  non-Abelian Alice strings of different color magnetic fluxes with
  total color flux canceled out are joined at a junction called a
  Boojum.
\end{abstract}
\maketitle

\section{Introduction}

Neutron stars, particularly pulsars, provide us with a unique
opportunity to study states of matter under extreme
conditions: highest known baryon density in the universe, rapid
rotation, strong magnetic fields, etc (see, e.g.,
Refs.~\cite{Baym:2017whm, Graber:2016imq, Kojo:2020krb} for reviews).
In the present work, we address the combined effect of high-density
and rapid rotation; namely the quantum vortices that appear in
the neutron superfluid and the color-superconducting quark matter.

There are several ongoing observations of neutron stars
such as the ones with two-solar-mass~\cite{Demorest:2010bx,
  *Fonseca:2016tux, *Arzoumanian:2017puf, Antoniadis1233232,
  Cromartie:2019kug}, probed by gravitational wave
detectors~\cite{TheLIGOScientific:2017qsa, Abbott:2020uma}, and
investigated in the NICER mission~\cite{Riley:2019yda,
  Miller:2019cac}, etc., and they 
have led to a flurry of studies about neutron stars in diverse fields of
research.
Amongst such efforts, numerous articles focus on the equation of state
(EoS).
The recent development in this direction was constructing a
phenomenological hybrid EoS with a smooth crossover for the
hadron-to-quark phase transition~\cite{Masuda:2012kf,*Masuda:2012ed,
  Kojo:2014rca, Baym:2017whm, Baym:2019iky, Kojo:2020ztt} rather than
a first-order phase transition as conventionally done in many
literatures.
The crossover construction enables neutron stars to have a sizable
quark core inside consistently with the constraints put by
observations.
Such construction should not be regarded as an exotic alternative
because there can indeed be a possibility that a substantial quark
core inside a heavy neutron star is realized as suggested by the
model-independent analysis~\cite{Annala:2019puf}.
The plausible ground state of such cold dense quark matter is color
superconductor~\cite{Bailin:1983bm, Alford:2007xm} with various
patterns of the diquark pairing being known such as color-flavor
locked (CFL) phase~\cite{Alford:1998mk} in three-flavor symmetric
matter and two-flavor superconducting (2SC) phase~\cite{Alford:1997zt,
  Rapp:1997zu} in two-flavor symmetric matter.

The idea of crossover construction of the EoS can be traced back to
the concept of \emph{quark-hadron continuity} between the CFL phase
and the hyperon superfluid phase ~\cite{Schafer:1998ef} (see
also~\cite{Alford:1999pa, Fukushima:2004bj,  Hatsuda:2006ps,
  *Yamamoto:2007ah, Hatsuda:2008is, Schmitt:2010pf}).
The both color-superconducting and hadronic superfluid phases share
the same symmetry breaking patterns and low-lying excitations, so
these phases can be connected continuously.
The recent breakthrough is an extension to the case with rapid
rotation under which superfluid vortices are created in hadronic and
quark matter~\cite{Alford:2018mqj, Chatterjee:2018nxe,
  *Chatterjee:2019tbz, Cherman:2018jir, Hirono:2018fjr, *Hirono:2019oup,
  Cherman:2020hbe}.
In the CFL phase, superfluid vortices appear~\cite{Forbes:2001gj,
  Iida:2002ev}, but each superfluid vortex is unstable against a decay
into a set of more stable vortices~\cite{Nakano:2007dr,
  *Nakano:2008dc, Alford:2016dco}.
The most stable vortices are non-Abelian semi-superfluid vortices
carrying color magnetic fluxes and fractionally quantized 1/3
circulation of the Abelian superfluid
vortices~\cite{Balachandran:2005ev, Nakano:2007dr, *Nakano:2008dc,
  Eto:2009kg, *Eto:2009bh, *Eto:2009tr, Eto:2013hoa, Alford:2016dco}.
Thus, an important question was raised in Ref.~\cite{Alford:2018mqj}
how these non-Abelian vortices penetrate into hyperonic matter;
It was suggested in Ref.~\cite{Alford:2018mqj} that one non-Abelian
vortex in the CFL phase should be connected to one superfluid vortex
in the hyperon matter.
This work stimulated the discussion whether there is a discontinuity
between the vortices in two phases~\cite{Cherman:2018jir} or
not~\cite{Hirono:2018fjr, *Hirono:2019oup} from the viewpoint of
topological order based on a Wilson loop linking a vortex.

Aharonov-Bohm (AB) phases of quasi-particle encircling vortices
provide us with further insight into this problem by comparing these
in the hyperonic and CFL phases.
It was shown in Ref.~\cite{Chatterjee:2018nxe, *Chatterjee:2019tbz}
that three non-Abelian vortices, which carry different color magnetic
fluxes with total color canceled out, must joint at one point to three
integer vortices in the hyperonic matter.
Such a junction point of vortices is called a colorful
\emph{Boojum}~\cite{Cipriani:2012hr}\footnote{
See Refs.~\cite{Blaschke:1999fy,*Sedrakian:2008ay} for earlier works on a similar structure.}; 
originally, the similar
structures found in helium superfluids were named Boojums by
Mermin~\cite{Mermin1977},
\footnote{
  Originally the boojum is a particular variety of the fictional
  animal species called {\it snarks} created by Lewis Carroll in his
  nonsense poem ``The Hunting of the Snark.'' See~\cite{Mermin1990}
  for the story how Mermin made ``boojum''  an internationally accepted
  scientific term.
}
and have been predicted to occur in $^{3}$He
superfluids~\cite{volovik2003universe} in particular at the A-B phase
boundary~\cite{PhysRevLett.89.155301,Bradley2008}, liquid
crystals~\cite{Carlson:1988}, Bose-Einstein
condensates~\cite{Kasamatsu:2013lda}, and
quantum field theory~\cite{Eto:2006pg}.
Moreover, in neutron stars, Boojums were suggested to explain pulsar
glitch phenomena, which is a sudden speed-up in
rotation~\cite{Marmorini:2020zfp}.

Thus far, we have discussed quark-hadron continuity within the ideal
three-flavor symmetric setup in the limit of strange ($s$) quark mass
degenerate with up ($u$) and down ($d$) quark masses.
In the realistic setup, however, the hadronic matter is dominated by
neutrons, which are composed of two-flavor $u$ and $d$ valence quarks,
and the $s$-quark masses are heavy so that they do not participate in
condensation.
Neutrons are superfluids, for which pairing in $^3P_2$ channel are
responsible~\cite{Hoffberg:1970vqj, Tamagaki:1970ptp, *takatsukaPTP71,
  *takatsukaPTP72, *Takatsuka:1992ga, richardsonPRD72, Sauls:1978lna}
(see also Refs.~\cite{Mizushima:2016fbn, *Mizushima:2019spl,
  Yasui:2018tcr, *Yasui:2019unp} and references therein for recent
studies).
By contrast, the conventional 2SC phase does not exhibit the property
of superfluidity because $\mathrm{U(1)_B}$ symmetry remains unbroken in
this phase.
It was, however, proposed recently in Ref.~\cite{Fujimoto:2019sxg,
  *Fujimoto:2020cho} that the two-flavor color superconductor can be
superfluids if we take into account the $^3P_2$ pairing of $d$-quarks
in addition to the 2SC pairing, and this novel phase was named
2SC+$\dd$ phase.
The consideration of the superfluidity led to the two-flavor
counterpart of the quark-hadron continuity, i.e., the 2SC+$\dd$ phase
and $^3P_2$ neutron superfluid phase are continuously connected.
The two-flavor continuity is indeed consistent with the crossover
construction of the EoS and more natural in the sense that we use the
hadronic EoS dominated by nucleons in the most of the cases; see
Ref.~\cite{Kojo:2020ztt} for an explicit crossover construction of the
EoS within the two-flavor setup.

Then, a natural question arises that if Boojums are also present in
the two-flavor quark-hadron continuity.
The most stable vortices in the two-flavor dense QCD are
``\emph{non-Abelian} Alice strings''~\cite{Fujimoto:2020dsa}, which
are the non-Abelian counterpart of the so-called Alice
strings~\cite{Schwarz:1982ec, Alford:1990mk, *Alford:1990ur,
  *Alford:1992yx, Leonhardt:2000km, Chatterjee:2017jsi,
  *Chatterjee:2017hya, *Chatterjee:2019zwx, *Nitta:2020ggi}.
These vortices have $\rm U(1)_B$ fractional windings along with the
color-magnetic fluxes.
Three non-Abelian Alice strings are more energetically favorable
than a single superfluid vortex, so that the latter decays into the
former.
Quasi-particles winding around a non-Abelian Alice string pick up
nontrivial (color non-singlet) AB phases, unlike those of CFL vortices
with color singlet AB phases.

In this work, we show that
in the two-flavor quark-hadron continuity picture,
three Alice strings with red, blue, green color magnetic fluxes 
in two-flavor quark matter must 
join at a junction point to three integer vortices 
in $^3P_2$ neutron matter, forming a colorful Boojum.

\section{Two-flavor dense matter}
\label{sec:dense}
We give a brief synopsis of the two-flavor hadronic and quark
matter at high density~\cite{Fujimoto:2019sxg, *Fujimoto:2020cho}.

The hadronic phase that we consider 
is a neutron $^3 P_2$ superfluid 
with 
the order parameter operator~\cite{Hoffberg:1970vqj,
  Tamagaki:1970ptp, Takatsuka:1992ga}
\begin{align}
  \hat{A}^{ij} \, =\,  \hat{n}^T \calC \gamma^i \nabla^j \hat{n}\,,\label{eq:defA}
\end{align}
with a neutron field operator $\hat{n}$  
and the charge conjugation operator $\calC$. 
Here, the Roman letters ($i,j,\ldots$) denote spatial
coordinates, and the matrices $\gamma^i$ and spatial derivatives
$\nabla^j$ account for spin and angular momentum contributions in the
$^3 P_2$ pairing, respectively.
We assume that neutrons made out of $u$- and $d$-quarks can be
described as a quark-diquark system $\hat{n} = \epsilon^{\alpha\beta\gamma}
(\hat{u}_\alpha^T \calC \gamma^5 \hat{d}_\beta) \hat{d}_\gamma$,
with 
the Greek letters ($\alpha,\beta,\ldots$) denoting color indices.

On the other hand, 
the relevant quark matter is the 2SC+$\dd$ phase, 
where the
expectation value of $\hat{A}^{ij}$ can be taken, within the mean field
approximation, as
\begin{align}
 & A^{ij} =  \langle \hat{A}^{ij} \rangle \simeq
  (\phiud)^\alpha (\phiud)^\beta
  (\phidd)_{\alpha\beta}^{ij}\,,
  \label{eq:aij}\\
  & \phiud \equiv
   \langle
   \epsilon^{\alpha\beta\gamma} 
  \hat{u}^T_{\beta} \calC \gamma^5 \hat{d}_{\gamma} \rangle
  , \quad
  \phidd \equiv 
  \langle \hat{d}^T_{\alpha} \calC
  \gamma^i \nabla^j \hat{d}_\beta \rangle. 
  \label{eq:dd}
\end{align}
Here, the spatial indices $i,j$ of $\phidd$ are suppressed, and
hereafter an appropriate tensor structure is implied.
Note that quantities with (without) hat symbols denote 
operators (condensates). 
The two condensates $\phiud$ and $\phidd$ 
in
Eq.~\eqref{eq:aij} account for the color superconductivity of the
quark matter:  $\phiud$ is the so-called 2SC condensate, while the
novel feature here is represented by $\phidd$, which is the diquark
condensate of $d$-quarks in the $^3P_2$ channel.
The symmetry of QCD relevant to this work is $G_{\rm QCD} =
\mathrm{SU(3)}_{\rm C} \times \mathrm{U(1)}_{\rm B} 
\ni (U, e^{i \thetab})$ 
  acting on quark fields $\hat q$ as $\hat q \to  e^{i
  \thetab} U \hat q$ as a column vector belonging to the fundamental
representation $\boldsymbol{3}$ of $\mathrm{SU(3)_C}$, 
under which the diquark condensates~(\ref{eq:dd}) transform as
\begin{equation}
  \phiud \to e^{2i\theta_{\rm B}} U^* \phiud\,,\quad
  \phidd \to e^{2i\theta_{\rm B}} U \phidd U^T\,.
  \label{eq:transf}
\end{equation}
Since $A^{ij}$ is non-zero in the both phases, 
the local order parameters indeed cannot distinguish 
these two phases, implying the quark-hadron 
continuity~\cite{Fujimoto:2019sxg, *Fujimoto:2020cho}.

\section{Vortices in two-flavor dense matter}
\label{sec:vortex}
We classify the vortices that appear in the two-flavor neutron and
quark matter. 
First, let us discuss $^3P_2$ neutron superfluid vortices.
In terms of the $^3P_2$ order parameter in Eq.~(\ref{eq:defA}),
a single integer vortex 
behaves at large distance as~\cite{
richardsonPRD72,
Muzikar:1980as,Sauls:1982ie,Masuda:2015jka,*Chatterjee:2016tml,*Masaki:2019rsz}
\begin{align}
  A^{ij} (\varphi) \sim e^{i\varphi} A^{ij}(\varphi=0).
  \label{eq:single-3P2}
\end{align}
In the weak coupling limit, the $^3P_2$ superfluid is in the nematic
phase~\cite{Sauls:1978lna} with $A^{ij}(\varphi=0) \sim \diag(s,s,1-s)$
with a real parameter $s$, which is actually determined by the
temperature and magnetic
field~\cite{Mizushima:2016fbn,Yasui:2018tcr,Yasui:2019unp}.
At higher magnetic field, half-quantized vortices are the most
stable~\cite{Masuda:2016vak}, but we do not consider such a case for
simplicity.

Next, let us discuss vortices in the 2SC+$\dd$
phase~\cite{Fujimoto:2020dsa}.
The simplest vortex is what 
we call a $\rm U(1)_B$ superfluid or an Abelian vortex, given by 
\begin{align}
  \phidd(\varphi) &= f_0(r) e^{i \varphi} \deltadd{\bf 1}_3 \sim e^{i
    \varphi} \deltadd{\bf 1}_3\,\nonumber \\
  \Phi_{\rm  2SC}   &=  
  h(r) \frac{e^{i\varphi}}{\sqrt{3}} \Delta_{\rm  2SC} (1,1,1)^T
  \label{eq:Abelian}
\end{align}
with the boundary condition $f_0(0)=h(0)=0$ and $f_0(\infty)=h(\infty)=1$ 
for the profile functions $f_0(r)$ and $h(r)$, and $(r, \varphi)$ being
the polar coordinates.  We have also assumed a unitary gauge fixing for the
diquark condensates.
This vortex carries a unit quantized circulation in
$\mathrm{U(1)_B}$ as encoded in the factor $e^{i\varphi}$,
and thus is created under rotation because of the superfluidity.
This is topologically stable due to
$\pi_1[\mathrm{U(1)_B}] = \mathbb{Z}$ but is dynamically unstable against decay
into three non-Abelian Alice strings introduced below.  The $^3P_2$
order parameter in  Eq.~(\ref{eq:aij}) behaves at large distance as
\begin{align}
  A^{ij} \sim e^{3 i\varphi} A^{ij}(\varphi=0) ,
\end{align}
corresponding to three
integer vortices in the $^3P_2$ neutron matter,
compared with Eq.~\eqref{eq:single-3P2}

Next, we present a non-Abelian Alice string as the most stable vortex,
behaving at spatial infinity as 
\begin{align}
  \phidd(\varphi) &\sim  e^{i\varphi/3} U(\varphi) \phidd(\varphi=0) U^T(\varphi)\,,\\
  U(\varphi) &= \mathcal{P} \exp\left(i g \int_0^\varphi
  \boldsymbol{A}\cdot d\boldell\right)\,.
  \label{eq:holonomy}
\end{align}
The full vortex ansatz is of the form of
\begin{equation}
  \begin{split}
    \phidd (\varphi)
    &= \deltadd
    \diag ( f(r) e^{i\varphi} ,g(r),g(r))
\,,\\
    U(\varphi) &= e^{i(\varphi/6)\diag(2,-1,-1)}\,,\\
    A_i &=- \frac{a(r)}{6 g}\frac{\epsilon_{ij} x^j}{r^2} 
    \diag(2,-1,-1)\,, \\
\Phi_{\rm  2SC}   &= (\Delta_{\rm  2SC}  ,0,0)^T,
  \end{split}
  \label{eq:ansatz3}
\end{equation}
for the red color magnetic flux ($r$), 
\begin{equation}
  \begin{split}
    \phidd (\varphi)
    &= \deltadd
        \diag ( g(r),f(r) e^{i\varphi},g(r))
\,,\\
    U(\varphi) &= e^{i(\varphi/6)\diag(-1,2,-1)}\,,\\
    A_i &= - \frac{a(r)}{6 g}\frac{\epsilon_{ij} x^j}{r^2} 
    \diag(-1,2,-1)\,,\\
    \Phi_{\rm  2SC}   &= (0,\Delta_{\rm  2SC} ,0)^T,
  \end{split}
  \label{eq:ansatz2}
\end{equation}
for the green one ($g$), and 
\begin{equation}
  \begin{split}
  \phidd (\varphi)
  &= \deltadd
      \diag (g(r),g(r), f(r) e^{i\varphi} )
\,,\\
     U(\varphi) &= e^{i(\varphi/6)\diag(-1,-1,2)} \,,\\
  A_i &= - \frac{a(r)}{6 g}\frac{\epsilon_{ij} x^j}{r^2} 
  \diag(-1,-1,2)\,,\\
  \Phi_{\rm  2SC}   &= (0,0,\Delta_{\rm  2SC})^T,
  \end{split}
  \label{eq:ansatz}
\end{equation}
for the  blue one ($b$), 
with the boundary conditions for the profiles 
$f (0) = g' (0) =a(0) = 0, \quad f (\infty) =g(\infty) = a(\infty) = 1$.
These carry $1/6$ quantized color-magnetic fluxes, 
$\calF = \calF_0 / 6$, 
and $1/3$ quantized circulations.

For all the three cases, the $^3P_2$ order parameter given in
Eq.~(\ref{eq:aij}) behaves at large distance as
\begin{align}
  A^{ij} (\varphi) \sim e^{i\varphi} A^{ij}(\varphi=0),
  \label{eq:single-NA}
\end{align}
coinciding with Eq.~\eqref{eq:single-3P2}  of one 
integer vortex 
in the $^3P_2$ neutron matter.

\section{Vortex continuity and Boojums}
As mentioned earlier, the idea of the vortex continuity was originally
proposed in Ref.~\cite{Alford:2018mqj}.
Their discussion of the vortex continuity was based on quantity called
the Onsager-Feynman circulation whose definition is given by $C =
\oint \bv \cdot d\boldell = 2\pi n /\mu$ with $n$ and $\mu$ being the
winding number and chemical potential of the condensate, respectively.
The circulations of vortices in the hadronic and quark phase are turned
out to be identical, so, it led to the observation that a single
hadronic vortex would be smoothly connected to a single non-Abelian
vortex.
One can also calculate the circulation for our case:  The circulation
of a neutron $^3P_2$ vortex is $C_{nn} = \pi / \mu_{\rm B}$ with
$\mu_{\rm B}$ being the baryon chemical potential.  The circulation of
a non-Abelian Alice string is given by $C_{\text{NA}} = \pi /3
\mu_{\rm q} = \pi / \mu_{\rm B}$, where $\mu_{\rm q} = \mu_{\rm B}/3$
is the quark chemical potential.  The circulations in the both
phases coincide with each other, $C_{nn} = C_{\text{NA}}$.
Equivalently, 
the expressions for the neutron superfluid order parameter in
Eqs.~(\ref{eq:single-3P2}, \ref{eq:single-NA}) coincide with each
other. Thus, at a glance one might think that a single non-Abelian
Alice string would be connected to a single integer $^3P_2$ vortex.
It is, however, \emph{not} true as shown below.

In order to investigate how vortices are connected, we employ the
(generalized) AB phases of quarks encircling around vortices. 
When the neutron field $\hat{n}$ encircles around a single integer
vortex given in Eq.~\eqref{eq:single-3P2}, it receives a phase factor
(generalized AB phase) $\exp (i\varphi/2)$ at angle $\varphi$, and
after the complete encirclement it obtains $\exp(i\pi)= -1$.
Then, we next take quarks as probe particles and calculate the AB
phase that the quarks receive.  Since we have assumed the neutron
operator as $\hat{n} = \epsilon^{\alpha\beta\gamma} (\hat{u}_\alpha^T
\calC \gamma^5 \hat{d}_\beta) \hat{d}_\gamma$, light quarks $\qop =
\hat{u}, \hat{d}$ obtain the phase $\Gamma_{nn}^{u,d} (\varphi) \equiv
\exp(i\varphi/6)$ when they encircle the neutron vortex at
angle $\varphi$:
\begin{equation}
  \qop(\varphi=0)\ \to\  \qop(\varphi)\sim
  \Gamma_{nn}^{u,d}(\varphi)\qop(\varphi=0)\,.
\end{equation}
Thus, at the quark level, the generalized AB phase forms a ${\mathbb
  Z}_6$ group.  The heavy quark field $\hat{s}$ receives no phase
around the vortex, i.e., $\Gamma_{nn}^{s}(\varphi) \equiv 1$.  We
explain our notation that $\Gamma_{nn}^{\psi}$ is the generalized AB
phase around neutron $^3P_2$ vortex probed by particle $\psi=\hat{u},
\hat{d}, \hat{s}$.

Let us turn to quark matter and calculate phase factors of quarks
around the Alice string.
For any $\varphi \neq 0$, the light quark field ($\qop=\hat{u},
\hat{d}$)
and heavy quark field ($\hat{s}$) are given by a holonomy action as
\begin{align}
  \qop(\varphi) &\sim e^{i\thetab(\varphi)} U(\varphi) \qop(\varphi = 0)\,, \\
  \hat s (\varphi) &\sim U(\varphi) \hat s (\varphi = 0)\,, 
\end{align}
respectively, where $U(\varphi)$ is defined in
Eq.~\eqref{eq:holonomy}.
Here, we use the following shorthand notation for the generalized AB
phase.  The field $\hat\psi (= \qop, \hat s, \ldots)$ of color $\beta
(= r,g,b)$ encircling around the flux tubes of the $\alpha(=r,g,b)$
color magnetic flux acquires the AB phase denoted by
$\Gamma^{\psi}_{\alpha\beta}$:
\begin{align}
  \hat\psi_{\beta}(\varphi=0) \to \hat\psi_{\beta}(\varphi) \sim \Gamma^{\psi}_{\alpha\beta}(\varphi) \hat\psi_{\beta}(\varphi=0)\,,
\end{align}
where the index $\beta$ is not contracted. 
The AB phases of a heavy quark encircling flux tubes are
\begin{align}
  \Gamma_{\alpha\beta}^{s} (\varphi)=
  \bordermatrix{ & r & g & b \cr
    r & e^{+i \varphi/3 } & e^{-i \varphi/6 } & e^{-i \varphi/6 } \cr
    g & e^{-i \varphi/6 } & e^{+i \varphi/3 } & e^{-i \varphi/6 } \cr
    b & e^{-i \varphi/6 } & e^{-i \varphi/6 } & e^{+i \varphi/3 }} \,,
\end{align}
where, as explicitly indicated above, the row ($\alpha=r,g,b$) denotes
the color of the flux tubes, and the column ($\beta=r,g,b$) denotes the
colors of the heavy ($s$) quark encircling them.
Thus, the heavy quark field $s$ forms a ${\mathbb Z}_6$ group around 
the Alice string.

When the light quarks $u,d$ encircle the Alice string, they also
receive $\rm U(1)_B$ transformation $e^{+i\varphi/6}$ as well as the
AB phase that they have in common with those of the $s$-quarks.  
Therefore, the generalized AB phases are
\begin{align}
  \Gamma_{\alpha\beta}^{u,d} (\varphi)
  = e^{i\varphi/6} \Gamma_{\alpha\beta}^{s} (\varphi) 
  =
  \begin{pmatrix}
    e^{i \varphi/2 } & 1 & 1   \\
    1 & e^{i \varphi/2 } & 1 \\
    1 & 1 & e^{i \varphi/2 }   \\
  \end{pmatrix}\,.
\end{align}  
Thus, the light quarks $u,d$ form a ${\mathbb Z}_2$ group around 
the Alice string. 

From the above calculations of generalized AB phases around the
vortices, one can immediately conclude that one integer vortex in
the $^3P_2$ phase cannot be connected to one Alice string with any color
flux.
We can check this by the fact that the AB phases of the light quarks
do not match between the hadron and two-flavor quark matters:
\begin{align}
  \Gamma_{nn}^{u,d}(\varphi) \neq \Gamma_{\alpha\beta}^{u,d}(\varphi) \mbox{ for any } \alpha,\beta=r,g,b
\end{align}
and the same for heavy quarks:
\begin{align}
  \Gamma_{nn}^{s} (\varphi) \neq \Gamma_{\alpha\beta}^{s}(\varphi) \mbox{ for any } \alpha,\beta=r,g,b.
\end{align}

\begin{figure}
  \centering
  \includegraphics[width=.9\linewidth,keepaspectratio]{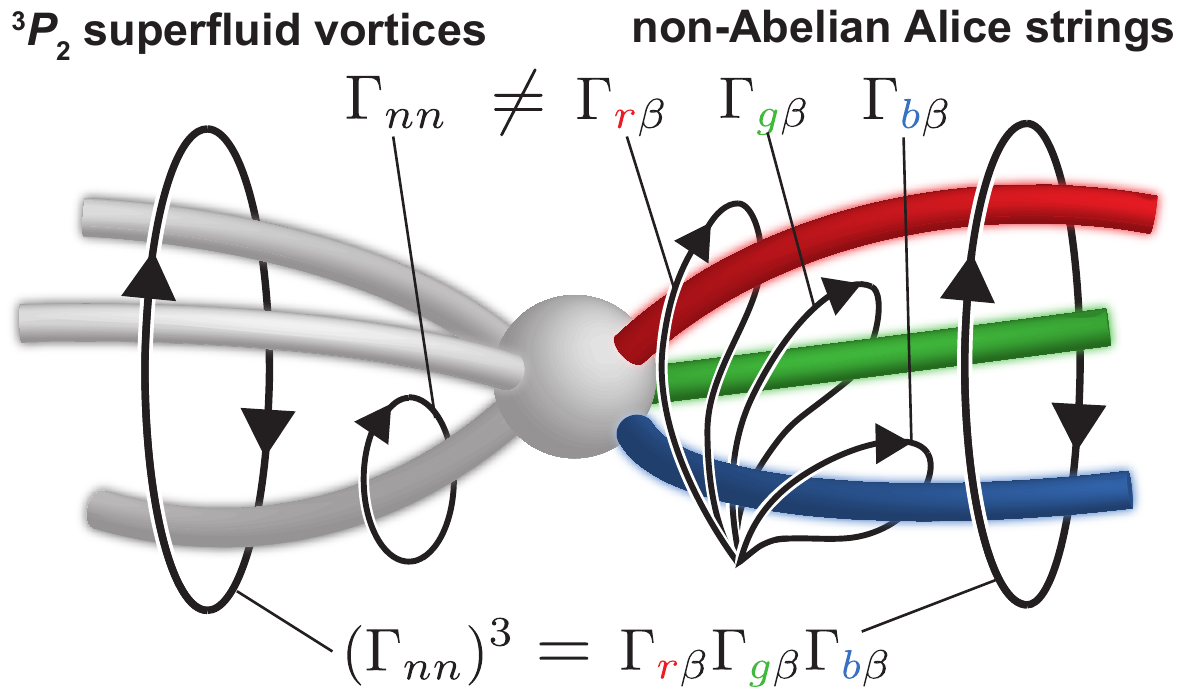}
  \caption{A schematic figure of the Boojum. Three $^3P_2$ neutron
    vortices in the hadronic phase are joined to three non-Abelian
    Alice strings in the color-superconducting phase.
    We also show the Aharonov-Bohm phase around each vortices.
    \label{fig:boojum} 
  }
\end{figure}

Only possibility is that a bundle of three integer vortices in the
$^3P_2$ neutron matter can be connected to a bundle of three Alice
strings with different color fluxes $r,g,b$.
In this case, the (generalized) AB phases for both phases completely
coincide:
\begin{align}
  [\Gamma_{nn}^{u,d}(\varphi)]^3
  = \Gamma_{r \beta}^{u,d} (\varphi)
  \Gamma_{g \beta}^{u,d} (\varphi)  
  \Gamma_{b \beta}^{u,d} (\varphi)
  \mbox{ for } \beta=r,g,b
\end{align}
for the light quarks $u,d$ of the color $\beta$, and
\begin{align}
  [\Gamma_{nn}^s(\varphi)]^3=
  \Gamma_{r \beta}^{s} (\varphi)
  \Gamma_{g \beta}^{s} (\varphi)  
  \Gamma_{b \beta}^{s} (\varphi)
  \mbox{ for } \beta=r,g,b
\end{align}
for the heavy quark $s$ of the color $\beta$.
We thus reach the picture of Boojum illustrated in
Fig.~\ref{fig:boojum}.

So far we have assumed that two-flavor dense QCD is in the so-called
deconfined phase in which non-Abelian Alice strings can exist.
On the other hand, if it is in the confined phase, non-Abelian Alice
strings must be confined either to doubly quantized non-Abelian
strings around which all AB phases are color singlet, or to U(1)$_{\rm
  B}$ Abelian strings.
In the context of quark-hadron continuity in Fig.~\ref{fig:boojum}, 
three non-Abelian Alice strings in the two-flavor quark matter 
is confined to one U(1)$_{\rm B}$ Abelian string. 
Thus, three integer vortices in the $^3P_2$ neutron superfluid
are combined to one Abelian string in  the two-flavor quark matter.

The mismatch between the AB phases of a
single hadronic vortex and a single non-Abelian vortex 
tells us that if we try to connect them, it might lead
to the discontinuity.  
The Boojums are therefore needed to maintain
the continuity.
Our discussion have been carried out on the level of the operator in
this work.  There can be a possibility that the AB phase may receive
the non-trivial contributions and the phase may be dynamically
screened if we take the expectation value under the vacuum, which may
lead to the different result, but we believe our present study already
captures the important feature and the further non-perturbative
analysis is the beyond the scope of this work.

\section{Summary}
In summary, we have found that the Boojums between the non-Abelian
Alice strings in the two-flavor quark matter (the 2SC+$\dd$
phase) and the $^3P_2$ neutron vortices in the hadronic matter.
As previously suggested in the three-flavor case, Boojum structures
are ubiquitous in quark-hadron continuity.
Neutron stars are rapidly rotating object so that they are accompanied
typically by about $10^{17}$ of vortices.  The vortices are believed
to play a crucial role in pulsar glitches, and it would be a
interesting and important question how our work influence such events.

The problem of the quark-hadron continuity is not only of the
phenomenological relevance, but also tightly related with the
fundamental problem of the gauge theory, particularly, the phase
structure of gauge theory with fundamental Higgs field. The idea that
the Higgs phase and the confinement phase is indistinguishable,
which is commonly referred to as Fradkin-Shenker
theorem~\cite{Osterwalder:1977pc, Fradkin:1978dv, Banks:1979fi}, is a
baseline for the idea of quark-hadron continuity, and is still under a
debate nowadays~\cite{Cherman:2020hbe}.
Since the non-perturbative studies of gauge theories are still limited
at present, we believe that the quark-hadron continuity in the bulk
matter and vortices from the neutron star phenomenology should serve
an important clues.

\begin{acknowledgments}
  We thank Shigehiro~Yasui for a discussion at the early stage of this
  work.
  This work is supported in part by Grant-in-Aid for Scientific
  Research, JSPS KAKENHI Grant Numbers 20J10506 (Y.F.) and JP18H01217
  (M.N.).
\end{acknowledgments}

\end{document}